\input{style/arxiv-ba.cfg}
\documentclass[ba,linksfromyear,preprint]{imsart}
\makeatletter
   \@ifpackageloaded{natbib}{}{\usepackage{natbib}}
\makeatother
\RequirePackage{amsthm,amsmath,amssymb}
\usepackage{algorithm}

\pubyear{2015}
\volume{10}
\issue{2}
\firstpage{297}
\lastpage{320}
\doi{10.1214/14-BA914}

\makeatletter
\setattribute{email}{text}{}

\newtheorem{theorem}{Theorem}[section]
\newtheorem{definition}{Definition}[section]

\renewcommand{\Re}{\mathbb{R}}
\providecommand{\by}{{\bf y}}

\DeclareMathOperator{\Be}{Beta}
\DeclareMathOperator{\Bino}{Binomial}
\DeclareMathOperator{\HG}{HG}
\DeclareMathOperator{\EHG}{EHG}

\def\Pr{\mathrm{Pr}}

\def\iidsamp{\; \raisebox{-0.5ex}{$\stackrel{\scriptscriptstyle{\text{iid}}}{\sim}$}\;}

\def\Oneup{\scriptscriptstyle(1)}
\def\Twoup{\scriptscriptstyle(2)}
\def\OneTwoup{\scriptscriptstyle(1,2)}

\def\Polya{P\'olya }
\makeatother

\begin{document}

\begin{frontmatter}
\title{Two-sample Bayesian Nonparametric Hypothesis Testing}
\runtitle{Two-sample BNP Hypothesis Testing}

\begin{aug}
\author[a]{\fnms{Chris C.} \snm{Holmes}\ead[label=e1]{cholmes@stats.ox.ac.uk}},
\author[b]{\fnms{Fran\c cois} \snm{Caron}\corref{}\ead[label=e2]{francois.caron@stats.ox.ac.uk}},
\author[c]{\fnms{Jim E.} \snm{Griffin}\ead[label=e3]{J.E.Griffin-28@kent.ac.uk}},
\and
\author[d]{\fnms{David A.} \snm{Stephens}\ead[label=e4]{d.stephens@math.mcgill.ca}}

\runauthor{C. C. Holmes et al.}


\address[a]{Department of Statistics and Oxford-Man Institute,
University of Oxford, England,\\\printead{e1}}
\address[b]{Department of Statistics, University of Oxford, England
\printead{e2}}
\address[c]{School of Mathematics, Statistics and Actuarial Science,
University of Kent, England,\\\printead{e3}}
\address[d]{Department of Mathematics and Statistics, McGill University,
Canada, \printead{e4}}

\end{aug}

%
\begin{abstract}
In this article we describe Bayesian nonparametric procedures for
two-sample hypothesis testing. Namely, given two sets of samples
$\mathbf{y}%
^{\Oneup}\iidsamp F^{\Oneup}$ and $\mathbf{y}^{\Twoup}\iidsamp
F^{\Twoup}$, with
$F^{\Oneup},F^{\Twoup}$ unknown, we wish to evaluate the evidence for
the null
hypothesis $H_{0}:F^{\Oneup}\equiv F^{\Twoup}$ versus the alternative
$H_{1}:F^{\Oneup}\neq F^{\Twoup}$. Our method is based upon a
nonparametric \Polya tree prior centered either subjectively or using
an empirical
procedure. We show that the \Polya tree prior leads to an analytic expression
for the marginal likelihood under the two hypotheses and hence an explicit
measure of the probability of the null $\Pr(H_{0}|\{ \mathbf
{y}^{\Oneup},
\mathbf{y}^{\Twoup} \} \mathbf{)}$.
\end{abstract}

%


\end{frontmatter}


\section{Introduction}

Nonparametric hypothesis testing is an important branch of statistics
with wide applicability. For example we often wish to evaluate the
evidence for
systematic differences between real-valued responses under two different
treatments without specifying an underlying distribution for the data.
That is, given two sets of samples $\mathbf{y}^{\Oneup}\iidsamp
F^{\Oneup}$ and
$\mathbf{y}^{\Twoup}\iidsamp F^{\Twoup}$, with $F^{\Oneup
},F^{\Twoup}$ unknown, we wish to
evaluate the evidence for the competing hypotheses
\[
H_{0}:F^{\Oneup}\equiv F^{\Twoup}\text{ versus }H_{1}:F^{\Oneup
}\neq F^{\Twoup}.
\]
In this article we describe a nonparametric Bayesian procedure for this
scenario. Our Bayesian method quantifies the weight of evidence
in favour of $H_{0}$ in terms of an explicit probability measure $\Pr
(H_{0}%
|\mathbf{y}^{\OneTwoup}\mathbf{)}$, where $\mathbf{y}^{\OneTwoup}$
denotes the
pooled data $\mathbf{y}^{\OneTwoup}=\break\{\mathbf{y}^{\Oneup},\mathbf{y}^{\Twoup}\}$.
To perform the test we use a \Polya tree prior
\citep{Lavine1992,Mauldin1992,Lavine1994} centered on some
distribution $G$ where under $H_0$ we have $F^{\OneTwoup}=F^{\Oneup
}=F^{\Twoup}$
and under $H_1$, $F^{\Oneup} \ne F^{\Twoup}$ are modelled as
independent draws from the \Polya tree prior. In this way we frame the
test as a model comparison problem and evaluate the Bayes Factor for
the two competing models. 
The \Polya tree is a well known nonparametric prior distribution for
random probability measures $F$ on $\Omega$ where $\Omega$ denotes
the domain of $Y$ \citep{Ferguson1974}.

Bayesian nonparametrics is a fast developing discipline, but while
there has been considerable interest in nonparametric inference there
has somewhat surprisingly been little written on nonparametric
hypothesis testing. Bayesian parametric hypothesis testing where
$F^{\Oneup}$ and $F^{\Twoup}$ are of known form is well developed in
the Bayesian literature, see e.g. \cite{Bernardo2000}, and most
nonparametric work has concentrated on testing a parametric model
versus a nonparametric alternative (the Goodness of Fit problem).
Initial work on the Goodness of Fit problem \citep{FlRiRo96,CaPa96}
used a Dirichlet process prior for the alternative distribution and
compared to a parametric model. In this case, the nonparametric
distributions will be almost surely discrete, and the Bayes factor will
include a penalty term for ties. The method can lead to misleading
results if the data is absolutely continuous, and has motivated the
development of methods using nonparametric priors that guarantee almost
surely continuous distributions. Dirichlet process mixture models are
one such class. The calculation of Bayes factors for Dirichlet
process-based models is discussed by \cite{basuchib}. Goodness of fit
testing using mixtures of triangular distributions is considered by
\cite{McVinish09}. An alternative form of prior, the \Polya tree, was
considered by \cite{Berger2001}. Simple conditions on the prior lead
to absolutely continuous distributions. \cite{Berger2001}
develop a default approach and consider its properties as a conditional
frequentist method. \cite{Hanson2006} discusses the use of Savage-Dickey
density ratios to calculate Bayes factors in favour of the centering
distribution (see also \cite{Branscum08}). Consistency issues are
discussed by \cite{DassLee04}, \cite{Rousseau2007}, \cite{Ghosal2008} and \cite{McVinish09}.
There has been some work on
testing the hypothesis that two distributions are the same; \cite{Dunson2008} consider hypothesis testing of stochastic ordering using
restricted Dirichlet process mixtures, but their methods could be
modified to allow two-sided hypotheses. They consider an interval null
hypothesis and rely on Gibbs sampling for posterior computation. \cite{PennellDunson2008} develop a Mixture of Dependent Dirichlet Processes
approach to testing changes in an ordered sequence of distributions
using a tolerance measure. \cite{Bhattacharya2012} develop an approach
for nonparametric Bayesian testing of differences between groups, with
the data within each group constrained to lie on a compact metric space
or Riemannian manifold.

Recently, following work presented here (originally posted on arXiv
in~\cite{Holmes2009}), \cite{Ma2011} propose to allow the two random
distributions under the alternative to randomly couple on different
parts of the sample space, thereby achieving borrowing of information.
Moreover, \cite{Chen2012} propose to use Lavine's (1992) partition for
each $F^{(j)}$ centered at the normal distribution. Their approach
enables generalization to more than two samples, but contrary to our
approach requires a truncation level to be set. They also follow \cite{Berger2001} by choosing the parameter $c$ that maximizes the Bayes
factor in favor of the alternative.

The rest of the paper is as follows. In Section 2 we discuss the \Polya
tree prior and derive the marginal probability distributions that
result from such a prior. In
Section 3 we describe our method and algorithm for calculating $\Pr
(H_{0}%
|\mathbf{y}^{\OneTwoup}\mathbf{)}$ based on a subjective partition.
In Section 4 we
discuss an empirical Bayes procedure where the \Polya tree priors are
centered on the empirical cdf of the joint data. Section 5 discusses
the sensibility of the procedures to tuning parameters. Section 6
provides a discussion of related approaches and Section 7 concludes
with a discussion of potential extensions.

\section{\Polya tree priors}

\begin{figure}
\includegraphics[scale=1.1]{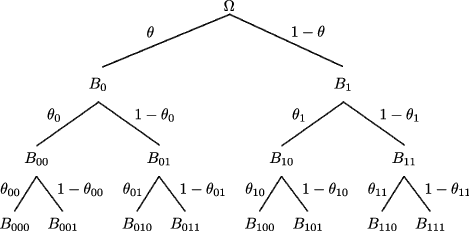}
\caption{Construction of a \Polya tree distribution. Each of the
$\theta_{\epsilon_m}$ is independently drawn from $\Be(\alpha
_{\epsilon_{m}0}, \alpha_{\epsilon_{m}1})$. Adapted from \cite
{Ferguson1974}.}
\label{fig:Polyatree}
\end{figure}

\Polya trees form a class of distributions for random probability
measures $F$ on some domain $\Omega$ \citep{Lavine1992,Mauldin1992,Lavine1994}.
Consider a recursive dyadic (binary) partition of $\Omega$ into
disjoint measurable sets. Denote the $k$th level of the partition
$\{B_{j}^{\scriptscriptstyle(k)},j=0,\ldots,2^k-1\}$, where
$B_{i}^{\scriptscriptstyle(k)}\cap B_{j}^{\scriptscriptstyle
(k)}=\varnothing$ for all $i\neq j$. The recursive partition is
constructed such that $B_{j}^{\scriptscriptstyle(k)} \equiv
B_{2j}^{\scriptscriptstyle(k+1)} \cup B_{2j+1}^{\scriptscriptstyle
(k+1)}$ for $k=1,2,\ldots, j=0,\ldots,2^k-1$. Figure~\ref
{fig:Polyatree} illustrates a bifurcating tree navigating the partition
down to level three for $\Omega=[0,1)$. It will be convenient to index
the partition elements using base 2 subscript and drop the superscript
so that, for example, $B_{000}$ indicates the first set in level 3,
$B_{0011}$ the fourth set in level 4 and so on.

To define a random measure on $\Omega$ we construct random measures on the
sets $B_{j}$. It is instructive to imagine a particle cascading down through
the tree such that at the $j$th junction the probability of turning
left or
right is $\theta_{j}$ and $(1-\theta_{j})$ respectively. In addition we
consider $\theta_{j}$ to be a random variable with some appropriate
distribution $\theta_{j}\sim\pi_{j}$. The sample path of the
particle down to
level $k$ will be recorded in a vector $\boldsymbol{\epsilon}_k=\{
\epsilon_{k1}%
,\epsilon_{k2},\ldots,\epsilon_{kk}\}$ with elements $\epsilon
_{ki}\in
\{0,1\}$, such that $\epsilon_{ki}=0$ if the particle went left at
level $i$,
$\epsilon_{ki}=1$ if it went right. Hence $B_{\boldsymbol{\epsilon
}_{k}}$ denotes
which partition the particle belongs to at level $k$. By convention,
set $\boldsymbol{\epsilon}_0=\emptyset$. Given a set of
$\theta_{j}$s it is clear that the probability of the particle falling
into the set $B_{\boldsymbol{\epsilon}_k}$ is just
\[
P(B_{\boldsymbol{\epsilon}_k})=\prod_{i=1}^{k}(\theta_{\boldsymbol
{\epsilon}_ {i-1}})^{(1-\epsilon_{ii} )}(1-\theta_{\boldsymbol
{\epsilon}_{i-1}})^{\epsilon_{ii}},
\]
which is just the product of the probabilities of falling left or right at
each junction that the particle passes through. This defines a random
measure on the partitioning sets.

Let $\Pi$ denote the collection of sets $\{B_{0},B_{1},B_{00},\ldots\}$
and let $\mathcal{A}$ denote the collection of parameters that
determine the
distribution at each junction, $\mathcal{A}=(\alpha_{00},\alpha
_{01},\alpha_{000},\ldots)$.

\begin{definition}{\cite{Lavine1992}}
A random probability measure $F$ on $\Omega$ is said to have a \Polya tree
distribution, or a \Polya tree prior, with parameters $(\Pi, \mathcal{A})$,
written $F \sim PT(\Pi,\mathcal{A})$, if there exists nonnegative numbers
$\mathcal{A} = (\alpha_{0}, \alpha_{1}, \alpha_{00}, \ldots)$ and
random variables ${\Theta} = (\theta,\theta_{0}, \theta_{1}, \theta
_{00}, \ldots)$ such that the following hold:

\begin{enumerate}
\item the random variables in ${\Theta}$ are mutually independent;

\item for every $k=1,2,\ldots$ and every $\boldsymbol{\epsilon}_k\in
\{0,1\}^k$,
\[
\theta_{\boldsymbol{\epsilon}_k}\sim\Be(\alpha
_{\boldsymbol{\epsilon}_k0},\alpha_{\boldsymbol{\epsilon}_k1});
\]

\item for every $k=1,2,\ldots$ and every $\boldsymbol{\epsilon}_k\in
\{0,1\}^k$,

\begin{equation}
F(B_{\boldsymbol{\epsilon}_k}|\Theta)=\prod_{i=1}^{k}(\theta
_{\boldsymbol{\epsilon}_{i-1}})%
^{(1-\epsilon_{ii})}(1-\theta_{\boldsymbol{\epsilon
}_{i-1}})^{\epsilon_{ii}}.\label{eq:PTproba}
\end{equation}
\end{enumerate}
\end{definition}

A random probability measure $F\sim PT(\Pi,\mathcal{A})$ is realized by
sampling the $\theta_{j}$s from the Beta distributions. $\Theta$ is
countably infinite as the tree extends indefinitely, and hence for most
practical applications the tree is specified only to a depth $m$. \cite
{Lavine1994} refers to this as a ``partially specified'' \Polya tree.
It is worth noting that we will not need to make this truncation in
what follows: our test will be fully specified with analytic
expressions for the marginal likelihood.\footnote{Note however that
consistency results only hold for a truncated version of the proposed test.}

By defining $\Pi$ and $\mathcal{A}$, the \Polya tree can be centered
on some chosen distribution $G$ so that $E [F] = G$ where $F\sim PT(\Pi,
\mathcal{A})$. Perhaps the simplest way to achieve this is to place the
partitions in $\Pi$ at the quantiles of $G$ and then set $\alpha
_{\boldsymbol{\epsilon}_k 0} = \alpha_{\boldsymbol{\epsilon}_j 1}$
for all $k=1,2,\ldots$ and all $\boldsymbol{\epsilon}_k\in\{0,1\}
^k$. \citep{Lavine1992}. For $\Omega\equiv\Re$ this leads to $B_{0}
= (-\infty, G^{-1}(0.5))$, $B_{1} =[G^{-1}(0.5), \infty)$ and, at
level $k$,
\begin{equation}
B_{\boldsymbol{\epsilon}_k} = [G^{-1}\{(k^{*}-1)/2^{k} \},
G^{-1}(k^{*}/2^{k})),
\end{equation}
where $k^{*}$ is the decimal representation of the binary number
$\boldsymbol{\epsilon}_k$.

It is usual to set the $\alpha$'s to be constant in a level $\alpha
_{\boldsymbol{\epsilon}_m0}=\alpha_{\boldsymbol{\epsilon
}_m1}=c_{m}$ for some constant $c_{m}$. The
setting of $c_{m}$ governs the underlying continuity of the resulting $F$'s.
For example, setting $c_{m}=c m^{2}$, $c > 0$, implies that $F$ is
absolutely continuous with probability 1 while $c_{m}=c/2^{m}$ defines
a Dirichlet process which makes $F$ discrete with probability 1 \citep{Lavine1992,Ferguson1974}. We will
follow the approach of \cite{Walker1999} and define $c_{m}=cm^{2}$. The choice of $c$ is discussed
in Section \ref{sec:sensitivity}.

\subsection{Conditioning and marginal likelihood}

An attractive feature of the \Polya tree prior is the ease with which
we can condition on data. \Polya trees exhibit conjugacy: given a
\Polya tree prior $F\sim PT(\Pi,\mathcal{A})$ and data $\mathbf{y}$
drawn independently from $F$, then \textit{a posteriori} $F$ also has
a \Polya tree distribution, $F|\mathbf{{y}\sim}PT\mathbf{(\Pi
,\mathcal{A}^{\ast})}$ where $\mathcal{A}^{\ast}$ is the set
of updated parameters, $\mathcal{A}^{\ast}=\{\alpha_{00}^{\ast
},\alpha
_{01}^{\ast},\alpha_{000}^{\ast},\ldots\}$
\begin{equation}
\alpha_{\boldsymbol{\epsilon}_i}^{\ast}|\mathbf{{y}=}\alpha
_{\boldsymbol{\epsilon}_i}+n_{\boldsymbol{\epsilon}_i}\mathbf{,}
\end{equation}
where $n_{\boldsymbol{\epsilon}_i}$ denotes the number of
observations in $\mathbf{y}$
that lie in the partition $B_{\boldsymbol{\epsilon}_i}$. The
corresponding random
variables $\theta_{j}^{\ast}$ are therefore distributed \emph{a posteriori}
as
\begin{equation}
\theta_{j}^{\ast}|\mathbf{{y}=}\Be(\alpha_{j0}+n_{j0},\alpha
_{j1}+n_{j1})\label{eq:ThetaPost}
\end{equation}
where $n_{j0}$ and $n_{j1}$ are the numbers of observations falling
left and
right at the junction in the tree indicated by $j$. This conjugacy
allows for a straightforward calculation of the marginal likelihood for
any set of observations, as
\begin{align}
\Pr(\mathbf{{y}|\Theta,\Pi,\mathcal{A})} & =\prod_{j}\theta
_{j}^{n_{j0}%
}(1-\theta_{j})^{n_{j1}}\label{eq:PTlikelihood}
\end{align}
where $\theta_{j}|\mathcal{A} \sim Be(\alpha_{j0},\alpha_{j1})$ and
where the product in~\eqref{eq:PTlikelihood} is over the set of all
partitions, $j\in
\{0,1,00,\ldots,\}$, though clearly for many partitions we have $n_{j0}
=n_{j1}=0$. Equation~\eqref{eq:PTlikelihood} has the form of a product
of independent
Binomial-Beta trials hence the marginal likelihood is,
\begin{equation}
\Pr(\mathbf{{y}|\Pi,\mathcal{A})=}\prod_{j}\left( \frac{\Gamma
(\alpha
_{j0}+\alpha_{j1})}{\Gamma(\alpha_{j0})\Gamma(\alpha_{j1})}\frac
{\Gamma
(\alpha_{j0}+n_{j0})\Gamma(\alpha_{j1}+n_{j1})}{\Gamma(\alpha_{j0}%
+n_{j0}+\alpha_{j1}+n_{j1})}\right)
\label{eq:PTmarglikelihood}
\end{equation}
where $j\in\{0,1,00,\ldots,\}$. This marginal probability will form
the basis
of our test for $H_{0}$ which we describe in the next section.

\section{A procedure for Bayesian nonparametric hypothesis testing}

We are interested in providing a weight of evidence in favour of
$H_{0}$ given
the observed data. From Bayes theorem,
\begin{equation}
\Pr(H_{0}|\mathbf{y}^{\OneTwoup} ) \; \propto\; \Pr\mathbf
{(}\mathbf{y}%
^{\OneTwoup}\mathbf{|}H_{0})\Pr(H_{0})\mathbf{.}%
\end{equation}
Under the null hypothesis $H_{0}$, $\mathbf{y}^{\Oneup}$ and $\mathbf
{y}^{\Twoup}$ are samples from some common distribution $F^{\OneTwoup
}$ with $F^{\OneTwoup}$ unknown. We specify our uncertainty in
$F^{\OneTwoup}$ via a \Polya tree prior, $F^{\OneTwoup}\sim PT(\Pi
,\mathcal{A})$. Under $H_{1}$, we assume $\mathbf{y}^{\Oneup}\sim
F^{\Oneup}$,
$\mathbf{y}^{\Twoup}\sim F^{\Twoup}$ with $F^{\Oneup},F^{\Twoup}$
unknown. Again we adopt a \Polya tree prior for $F^{\Oneup}$ and
$F^{\Twoup}$ with the same prior parameterization as for $F^{\OneTwoup
}$ so that
\begin{equation}
F^{\Oneup},F^{\Twoup},F^{\OneTwoup} \; \iidsamp\; PT(\Pi,\mathcal{A})
\end{equation}
The logic for adopting a common prior distribution is that we regard
the $F$s as random draws from some universe of distributions that we
describe probabilistically through the \Polya tree distribution. $\Pi$
is constructed from the quantiles of some {\em{a priori}} centering
distribution. Following the approach of \cite{Walker1999,Mallick2003}
we take common values for the $\alpha_{j}$s at each level as $\alpha
_{j0}=\alpha_{j1}=cm^{2}$ for an $\alpha$ parameter at level $m$.

The posterior odds on $H_0$ is
\begin{equation}
\frac{\Pr(H_{0}|\mathbf{y}^{\OneTwoup}\mathbf{)}}{\Pr
(H_{1}|\mathbf{y}^{\Oneup}%
\mathbf{,}\mathbf{y}^{\Twoup}\mathbf{)}}=\frac{\Pr(\mathbf
{y}^{\OneTwoup}%
\mathbf{|}H_{0}\mathbf{)}}{\Pr(\mathbf{y}^{\Oneup}\mathbf
{,}\mathbf{y}%
^{\Twoup}\mathbf{|}H_{1}\mathbf{)}}\frac{\Pr(H_{0})}{\Pr(H_{1})}%
\label{BF_formula}
\end{equation}
where the first term is just the ratio of marginal likelihoods, the
Bayes Factor, which from \eqref{eq:PTmarglikelihood} and conditional
on our specification of $\Pi$ and $\mathcal{A}$, is
\begin{align}
\frac{P(\mathbf{y}^{\OneTwoup}\mathbf{|}H_{0}\mathbf{)}}{P(\mathbf
{y}%
^{\Oneup}\mathbf{,}\mathbf{y}^{\Twoup}\mathbf{|}H_{1}\mathbf{)}}
& =\prod
_{j}b_j\label{eq:BF1}
\end{align}
where
\begin{align}
b_j=& \frac{\Gamma(\alpha_{j0})\Gamma(\alpha_{j1})}{\Gamma(\alpha
_{j0}+\alpha_{j1})}\frac{\Gamma(\alpha_{j0}+n_{j0}^{\Oneup
}+n_{j0}^{\Twoup})\Gamma(\alpha_{j1}+n_{j1}^{\Oneup}+n_{j1}^{\Twoup
})}{\Gamma(\alpha_{j0}%
+n_{j0}^{\Oneup}+n_{j0}^{\Twoup}+\alpha_{j1}+n_{j1}^{\Oneup
}+n_{j1}^{\Twoup})}
\nonumber\\
& \qquad\qquad \times\frac{\Gamma(\alpha_{j0}+n_{j0}^{\Oneup
}+\alpha
_{j1}+n_{j1}^{\Oneup})}{\Gamma(\alpha_{j0}+n_{j0}^{\Oneup})\Gamma
(\alpha_{j1}%
+n_{j1}^{\Oneup})}\frac{\Gamma(\alpha_{j0}+n_{j0}^{\Twoup}+\alpha
_{j1}+n_{j1}
^{\Twoup})}{\Gamma(\alpha_{j0}+n_{j0}^{\Twoup})\Gamma(\alpha
_{j1}+n_{j1}^{\Twoup})}
\end{align}
and the product in \eqref{eq:BF1} is over all partitions, $j\in\{
\emptyset,0,1,00,\ldots,\}$, $n_{j0}^{\Oneup}$ and $n_{j1}^{\Oneup
}$ represent the numbers of observations in $\mathbf{y}^{\Oneup}$
falling right and left at each junction and $n_{j0}^{\Twoup}$ and
$n_{j1}^{\Twoup}$ are the equivalent quantities for $\mathbf
{y}^{\Twoup}$. We can see from \eqref{eq:BF1} that the overall Bayes
Factor has the form of a product of Beta-Binomial tests at each
junction in the tree to be interpreted as ``does the data support one
$\theta_j$ or two, $\{\theta_j^{\Oneup}, \theta_j^{\Twoup}\}$, in
order to model the distribution of the observations going left and
right at each junction?'', where for each $j$, $\theta_j \sim\Be
(\alpha_j,\alpha_j)$. The product in \eqref{eq:BF1} is defined over
the infinite set of partitions. However, for each branch $j$, $b_j=1$
if $n_{j0}^{\Oneup}+n^{\Oneup}_{j1}=0$ or $n_{j0}^{\Twoup}+n^{\Twoup
}_{j1}=0$; hence to calculate~\eqref{eq:BF1} for the infinite
partition structure we just have to multiply terms from junctions which
contain at least some data from the two sets of samples. Hence, we only
need specify $\Pi$ to the quantile level where partitions contain
observations from both samples.
Note also that in the complete absence of data (that is, when
$n_{j0}^{\Oneup}+n_{j0}^{\Twoup} = n_{j1}^{\Oneup}+n_{j1}^{\Twoup}
= 0$)
\[
b_j= \frac{\Gamma(\alpha_{j0})\Gamma(\alpha_{j1})}{\Gamma(\alpha
_{j0}+\alpha_{j1})}\frac{\Gamma(\alpha_{j0})\Gamma(\alpha
_{j1})}{\Gamma(\alpha_{j0}
+\alpha_{j1})}\frac{\Gamma(\alpha_{j0}+\alpha
_{j1})}{\Gamma(\alpha_{j0})\Gamma(\alpha_{j1})}
\frac{\Gamma(\alpha_{j0}+\alpha_{j1})}{\Gamma(\alpha_{j0})\Gamma
(\alpha_{j1})} =1
\]
for all $j$, so the Bayes Factor is 1.

The test procedure is described in Algorithm~\ref{algo:test1}.
\begin{algorithm}
\caption{Bayesian nonparametric test}
\label{algo:test1}
\begin{enumerate}
\item Fix the binary tree on the quantiles of some centering
distribution $G$.

\item For level $m=1,2,\ldots,$ for each $j$ set $\alpha_{j}=c m^{2}$
for some $c$.

\item Add the log of the contributions of terms in~\eqref{eq:BF1} for
each junction in
the tree that has non-zero numbers of observations in $\mathbf
{y}^{\OneTwoup}$
going both right and left.

\item Report $\Pr(H_{0}|\mathbf{y}^{\OneTwoup}\mathbf{)}$ as
$ \Pr(H_{0}|\mathbf{y}^{\OneTwoup}\mathbf{)=}\frac{1}{1+\exp
(-LOR)}%
$, where $LOR$ denotes the log odd ratio calculated at step 3.
\end{enumerate}
\end{algorithm}

\subsection{Prior specification}

The Bayesian procedure requires the specification of $\{\Pi, \mathcal
{A}\}$ in
the \Polya tree. While there are good guidelines for setting $\mathcal
{A}$ the setting of $\Pi$ is more problem specific, and the results
will be quite sensitive to this choice. Our current, default, guideline
is to first standardise the joint data $\mathbf{y}^{\OneTwoup}$ with
the median and interquantile range of $\mathbf{y}^{\OneTwoup}$ and
then set the partition on the quantiles of a standard normal density,
$\Pi= \Phi(\cdot)^{-1}$. We have found this to work well as a
default in most situations, though of course the reader is encouraged
to set $\Pi$ according to their subjective beliefs.

In our algorithm, parameter $c$ is treated as a fixed hyperparameter.
As we demonstrate in Section \ref{sec:Consistency}, a truncated
version of our test with a subjective partition is consistent under
null and alternative hypotheses irrespective of the choice of $c$.
However, $c$ does have an impact on finite sample properties, that is,
the finite sample posterior probabilities. This is always the case for
Bayesian model selection/hypothesis testing based on the marginal
likelihood, which is effectively a measure of how well the prior
predicts the observed data, and not a feature restricted to our
nonparametric procedure. In Section~\ref{sec:sensitivity} we provide
some guidelines on the sensitivity of the testing procedure to the
value of this parameter, and discuss empirical Bayes estimation of $c$.

\subsection{Consistency}

\label{sec:Consistency}
Conditions for the consistency of the procedure under the null
hypothesis and alternative hypothesis are developed for a related test
based on a truncation of the Bayes factor (\ref{BF_formula}). Let
$n=n_\emptyset^{\Oneup}+n_\emptyset^{\Twoup}$ be the total sample
size for both samples. Let $l(\varepsilon)$ be the length of the
vector $\varepsilon$. This also indicates that $B_{\epsilon}$ forms
part of the partition at level $l(\varepsilon)$, and in our
construction there are $2^{l(\epsilon)}$ partition elements at level
$l(\epsilon)$.
We consider the test statistics based on the truncated Bayes factor
\begin{equation}
BF_{\kappa_0}=\prod_{\{j|l(j)\leq\kappa_0\}}b_{j}
\end{equation}
where $\kappa_0\in\mathbb N$ defines the level of truncation and can
be set arbitrarily large. We also consider a truncated version of the
hypothesis test:
\begin{align*}
&H_{0,\kappa_0}:\forall\epsilon| l(\epsilon)\leq\kappa
_0,~~F^{\Oneup}(B_\epsilon)= F^{\Twoup}(B_\epsilon)~~\\
&\text{ versus }\\
&H_{1,\kappa_0}:\exists\epsilon| l(\epsilon)\leq\kappa_0 \text{
and } F^{\Oneup}(B_\epsilon)\neq F^{\Twoup}(B_\epsilon).
\end{align*}

First, assume $H_{0,\kappa_0}$ is true and let $F_0$ denote the true
distribution.
To prove consistency under $H_{0,\kappa_0}$, it is sufficient to show that
\[
\lim_{n\rightarrow\infty}\log BF_{\kappa_0}=
\infty
\]
as $n\rightarrow\infty$ if both samples are drawn from the same distribution.

\begin{theorem}
Suppose that the limiting proportion of observations in the first
sample exists and is $\beta_\emptyset$:
\begin{equation}
\beta_\emptyset=\lim_{n\rightarrow\infty}\frac{n_{\emptyset
}^{\Oneup}}{n_{\emptyset}^{\Oneup}+n_{\emptyset}^{\Twoup}}.
\label{eq:betaempty}
\end{equation}
If $0<\beta_\emptyset<1$
then, under $H_{0,\kappa_0}$,
\[
\lim_{n\rightarrow\infty} \log BF_{\kappa_0} =\infty
\]
and the test defined by Algorithm \ref{algo:test1}, truncated at level
$\kappa_0$, is consistent under the null.
\end{theorem}
\noindent\textbf{Proof.} See Appendix.
\ \rule{0.5em}{0.5em}

We now consider consistency under $H_{1,\kappa_0}$ for the truncated
version of the test.
\begin{theorem}
\label{theo:Bayesalter}
Assume that $0<\beta_\emptyset<1$, and that exists $B_\epsilon$,
$l(\epsilon)\leq\kappa_0$, such that $F^{\Oneup}(B_\epsilon)F^{\Twoup}(B_\epsilon)>0$
and $\frac{F^{\Oneup}(B_{\epsilon 0})}{F^{\Oneup}(B_\epsilon)}\neq\frac{F^{\Twoup}(B_{\epsilon 0})}{F^{\Twoup}(B_\epsilon)}$.
Then
\[
\lim_{n\rightarrow\infty} BF_{\kappa_0} =0
\]
and the test defined by Algorithm \ref{algo:test1}, truncated at level
$\kappa_0$, is consistent under the alternative.
\end{theorem}
\noindent\textbf{Proof.} See Appendix.
\ \rule{0.5em}{0.5em}

The proofs for consistency for the non-truncated test are much more
challenging, as one needs to bound terms at each level of the \Polya
tree. In the next section, we provide numerical experiments on the
evolution of the Bayes factor with respect to the sample size under
both $H_0$ and $H_1$, suggesting consistency for the non-truncated test.

\subsection{Simulations}

To examine the operating performance of the method we consider the following
experiments designed to explore various canonical departures from the null.

\begin{figure}
\includegraphics{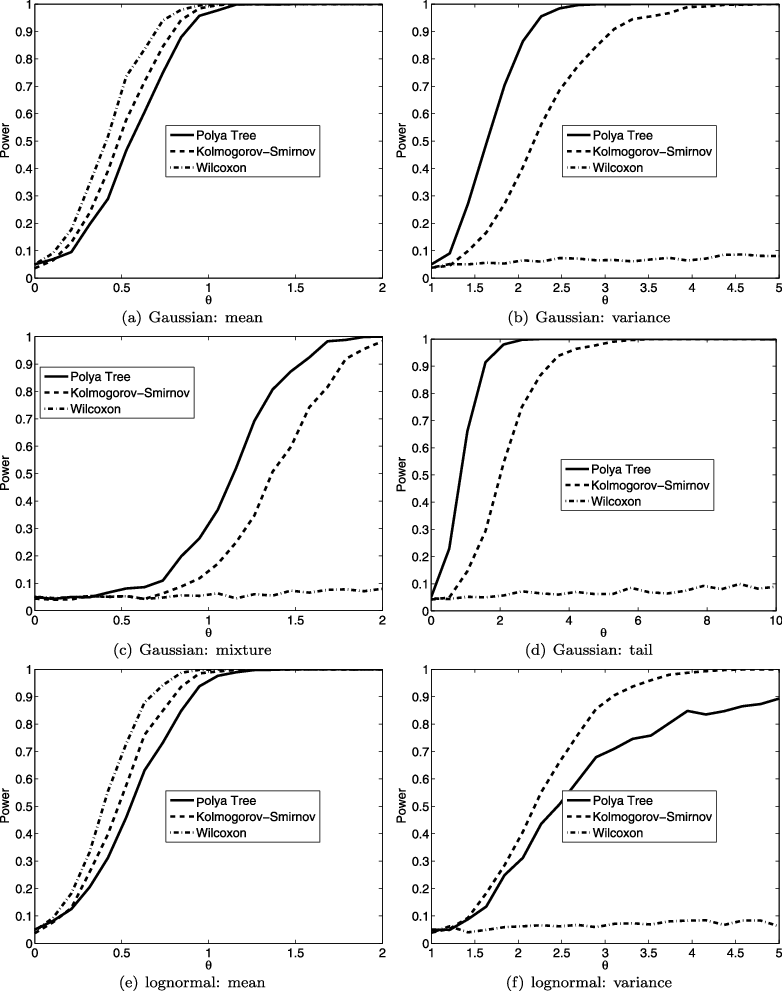}
\caption{Power of Bayes test with $\alpha_j=m^2$ on simulations from
Section 3.3., with
x-axis measuring $\theta$, the parameter in the alternative. Legend:
K-S (dashed), Wilcoxon (dot-dashed), Bayesian test (solid).}
\label{fig:fig1}
\end{figure}

\begin{itemize}
\item[a)] Mean shift: $Y^{\Oneup}\sim\mathcal{N}(0,1)$, $Y^{\Twoup
}\sim\mathcal{N}(\theta,1)$, $\theta=0,\ldots,3$.
\item[b)] Variance shift: $Y^{\Oneup}\sim\mathcal{N}(0,1)$,
$Y^{\Twoup}\sim\mathcal{N}(0,\theta^{2})$, $\theta=1,\ldots,3$.
\item[c)] Mixture: $Y^{\Oneup}\sim\mathcal{N}(0,1)$, $Y^{\Twoup
}\sim\frac{1}{2}\mathcal{N}(\theta,1)+\frac{1}{2}\mathcal
{N}(-\theta,1)$, $\theta=0,\ldots,3$.
\item[d)] Tails: $Y^{\Oneup}\sim\mathcal{N}(0,1)$, $Y^{\Twoup}\sim
t(\theta^{-1})$, $\theta=10^{-3},\ldots,10$ where $t(\nu)$ denotes
the standard Student t distribution with $\nu$ degrees of freedom.
\item[e)] Lognormal mean shift: $\log Y^{\Oneup}\sim\mathcal
{N}(0,1)$, $\log Y^{\Twoup}\sim\mathcal{N}(\theta,1)$, $\theta
=0,\ldots,3$.
\item[f)] Lognormal variance shift: $\log Y^{\Oneup}\sim\mathcal
{N}(0,1)$, $\log Y^{\Twoup}\sim\mathcal{N}(0,\theta^{2})$, $\theta
=1,\ldots,3$.
\end{itemize}
The default mean distribution $F_0^{\OneTwoup}= \mathcal{N}(0,1)$ was
used in the \Polya tree to construct the partition $\Pi$ and $\alpha
=m^2$. Data are standardized. Comparisons are performed with
$n_0=n_1=50$ against the two-sample Kolmogorov-Smirnov and Wilcoxon
rank test.
To compare the models we explore the ``power to detect the
alternative''. As a
test statistic for the Bayesian model we simulate data under the null
and then
take the empirical $0.95$ quantile of the distribution of Bayes Factors
as a
threshold to declare $H_1$. This is known as ``the Bayes, non-Bayes
compromise'' by~\cite{Good1992}. Results, based on 1000 replications,
are reported in Figure~\ref{fig:fig1}.
As a general rule we can see that the KS test is more sensitive to
changes in
central location while the Bayes test is more sensitive to changes to
tails or
higher moments.

\begin{figure}
\includegraphics{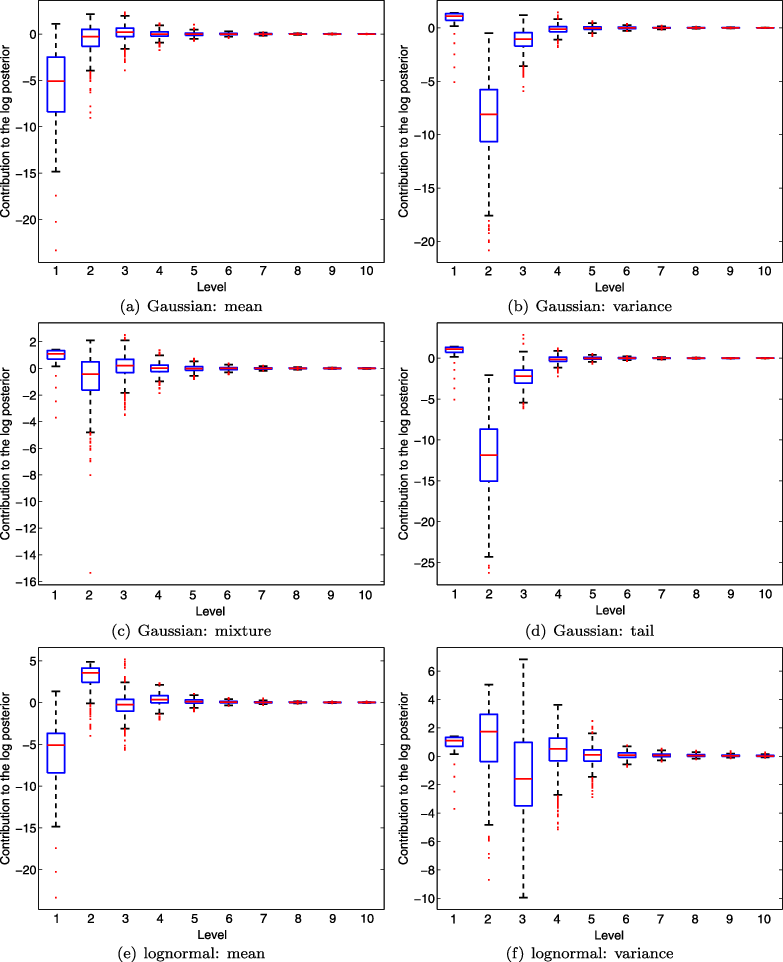}
\caption{Contribution to Bayes Factors from different levels of the
\Polya Tree under the alternative. Gaussian distribution with varying
(a) mean (b) variance (c) mixture (d) tails; log-normal distribution
with varying (e) mean (f) variance, from Section 3.2. Parameters of
$H_1$ were set to the mid-points of the x-axis in Figure~\ref{fig:fig1}.}
\label{fig:fig3}
\end{figure}

The dyadic partition structure of the \Polya Tree allows us to
breakdown the
contribution to the Bayes Factor by levels. That is, we can explore the
contribution, in the log of equation (10), by level. This is shown in
Figure~\ref{fig:fig3} as boxplots of the distribution of log BF statistics
across the levels for the simulations generated for Figure~\ref
{fig:fig1}. This
is a strength of the \Polya tree test in that it provides a qualitative and
quantitative decomposition of the contribution to the evidence against
the null
from differing levels of the tree.

\begin{figure}
\includegraphics{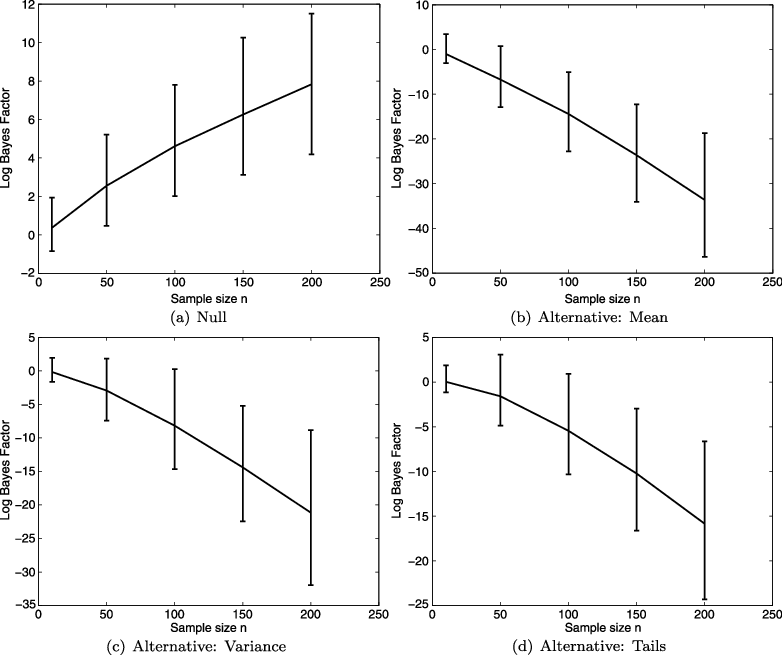}
\caption{Mean Bayes factor and 90\% confidence interval with respect
to the sample size $n$ under (a) the null and (b-c) two Gaussian
distributions with different (b) means, (c) variances and (d) a
Gaussian and a Student t.}
\label{fig:consist}
\end{figure}

It is also of interest to investigate the behavior of the Bayes factor
as a function of the sample size, both under the null and various
alternatives. Under the alternative, we consider in particular the
following cases:
\begin{enumerate}
\item[a)] Mean shift: $Y^{\Oneup}\sim\mathcal{N}(0,1)$, $Y^{\Twoup
}\sim\mathcal{N}(1,1)$.
\item[b)] Variance shift: $Y^{\Oneup}\sim\mathcal{N}(0,1)$,
$Y^{\Twoup}\sim\mathcal{N}(0,4)$.
\item[c)] Tails: $Y^{\Oneup}\sim\mathcal{N}(0,1)$, $Y^{\Twoup}\sim
t(1)$.
\end{enumerate}
The results are reported in Figure~\ref{fig:consist} for sample size
$n=10,50,100,200$ with 500 replications, and seem to suggest that the
non-truncated test is consistent under the null and alternative.

\section{A conditional procedure}

The Bayesian procedure above requires the subjective specification of the
partition structure $\Pi$. This subjective setting may make some users uneasy
regarding the sensitivity to specification.
In this section we explore an empirical procedure whereby the partition
$\Pi$ is centered on the data via the empirical cdf of the joint data
$\widehat{\Pi} = [ \widehat{F}^{\OneTwoup} ]^{-1}$.

Let $\widehat{\Pi}$ be the partition constructed with the quantiles
of the
empirical distribution $\widehat{F}^{\OneTwoup}$ of $\by^{\OneTwoup
}$. Under $H_{0}$, there are now no free parameters and only one degree
of freedom in the random variables $\{n_{j0}^{\Oneup}, n_{j1}^{\Oneup
}, n_{j0}^{\Twoup}, n_{j1}^{\Twoup}\}$ as conditional on the
partition centered on the empirical cdf of the joint, once one of the
variables has been specified the others are then known. We consider,
arbitrarily, the marginal distribution of $\{n_{j0}^{\Oneup}\}$\vadjust{\eject} which
is now a product of hypergeometric distributions (we only consider
levels where $n_{j}^{\OneTwoup}>1$)
\begin{align}
\Pr(\{n_{j0}^{\Oneup}\}|H_{0},\widehat{\Pi},\mathcal{A}) & \propto
\prod_{j}\left(
\begin{array}
[c]{c}%
n_{j}^{\Oneup}\\
n_{j0}^{\Oneup}%
\end{array}
\right) \left(
\begin{array}
[c]{c}%
n_{j}^{\OneTwoup}-n_{j}^{\Oneup}\\
n_{j0}^{\OneTwoup}-n_{j0}^{\Oneup}%
\end{array}
\right) \bigg/ \left(
\begin{array}
[c]{c}%
n_{j}^{\OneTwoup}\\
n_{j0}^{\OneTwoup}%
\end{array}
\right) \label{eq:H0}\\
& =\prod_{j}\HG\left(n_{j0}^{\Oneup};n_{j}^{\OneTwoup
},n_{j}^{\Oneup},n_{j0}^{\OneTwoup}\right)
\end{align}
if $\max(0,n_{j0}^{\OneTwoup}+n_{j}^{\Oneup}-n_{j}^{\OneTwoup})\leq
n_{j0}^{\Oneup}\leq\min(n_{j}^{\Oneup},n_{j0}^{\OneTwoup})$, and
zero otherwise.

Under $H_{1}$, the marginal distribution of $\{n_{j0}^{\Oneup}\}$ is a
product of the conditional distribution of independent binomial
variates, conditional on their sum,
\begin{equation}
\Pr\left(\left.\left\{n_{j0}^{\Oneup}\right\}\right
|H_{1},\widehat{\Pi},\mathcal{A}\right)\propto\prod_{j}\frac
{g\left(n_{j0}^{\Oneup};n_{j}^{\OneTwoup},n_{j}^{\Oneup
},n_{j0}^{\OneTwoup},\theta_{j}^{\Oneup}%
,\theta_{j}^{\Twoup}\right)}{\sum\limits_{x}g\left
(x;n_{j}^{\OneTwoup},n_{j}^{\Oneup},n_{j0}^{\OneTwoup}%
,\theta_{j}^{\Oneup},\theta_{j}^{\Twoup}\right)}\label{eq:H1}%
\end{equation}
if $\max(0,n_{j0}^{\OneTwoup}+n_{j}^{\Oneup}-n_{j}^{\OneTwoup})\leq
n_{j}^{\Oneup}\leq\min(n_{j}^{\Oneup},n_{j0}^{\OneTwoup} )$, zero
otherwise, and where
\begin{align*}
g\left(n_{j0}^{\Oneup};n_{j}^{\OneTwoup},n_{j}^{\Oneup
},n_{j0}^{\OneTwoup},\theta_{j}^{\Oneup}%
,\theta_{j}^{\Twoup}\right) & =\Bino\left(n_{j0}^{\Oneup
};n_{j}^{\Oneup},\theta_{j}^{\Oneup}\right) \times\ldots\\
& \Bino\left(n_{j0}^{\OneTwoup}-n_{j0}^{\Oneup};n_{j}^{\OneTwoup
}-n_{j}^{\Oneup},\theta
_{j}^{\Twoup}\right)
\end{align*}
and

\[
\theta_{j}^{\Oneup}|\mathcal{A} \sim\Be(\alpha_{j0},\alpha_{j1})
\qquad\qquad
\theta_{j}^{\Twoup}|\mathcal{A} \sim\Be(\alpha_{j0},\alpha_{j1}).
\]
Now, consider the odds ratio $\omega_{j}=\dfrac{\theta_{j}^{\Oneup
}(1-\theta_{j}^{\Twoup})}{\theta_{j}^{\Twoup}(1-\theta_{j}^{\Oneup})}$
and let
\[
\EHG\left(n_{j0}^{\Oneup};n_{j}^{\OneTwoup},n_{j}^{\Oneup
},n_{j0}^{\OneTwoup},\omega_{j}%
\right)=\frac{g(n_{j0}^{\Oneup};n_{j}^{\OneTwoup},n_{j}^{\Oneup
},n_{j0}^{\OneTwoup},\theta_{j}%
^{\Oneup},\theta_{j}^{\Twoup})}{\sum\limits
_{x}g(x;n_{j}^{\OneTwoup},n_{j}^{\Oneup},n_{j0}%
^{\OneTwoup},\theta_{j}^{\Oneup},\theta_{j}^{\Twoup})}.
\]
Then it can been seen that $\EHG(x;N,m,n,\omega)$ is the
extended hypergeometric distribution~\citep{Harkness1965} whose pdf is
proportional to
\[
\qquad\qquad\HG(x;N,m,n)\omega^x,\qquad\qquad a\leq x\leq b,
\]
where $a=\max(0,n+m-N)$, $b=\min(m,n)$. Note there are C++ and R
routines to evaluate the pdf. The extended hypergeometric distribution
models a biased urn sampling scheme whereby there is a different
likelihood of drawing one type of ball over another at each draw. The
Bayes factor is now given by
\begin{equation}
\label{eq:bfempi}
BF =\prod_{j}\frac{\HG\left(n_{j0}^{\Oneup};n_{j}^{\OneTwoup
},n_{j}^{\Oneup},n_{j0}^{\OneTwoup}%
\right)}{\displaystyle{\int_{0}^{\infty}} \EHG\left
(n_{j0}^{\Oneup};n_{j}^{\OneTwoup},n_{j}^{\Oneup},n_{j0}
^{\OneTwoup},\omega_{j}\right)p(\omega_{j})d\omega_{j}}%
\end{equation}
where the marginal likelihood in the denominator can be evaluated using
importance sampling or one-dimensional quadrature.

The conditional Bayes two-sample test can then be given in a similar
way to Algorithm~\ref{algo:test1} but now using (\ref{eq:bfempi}) for
the contribution at each junction.
Conditions for the consistency of the procedure under the null
hypothesis are developed for a related test based on a truncation of
the Bayes factor (\ref{eq:bfempi}) although we have been unable to
show consistency under the alternative.

\begin{theorem}
Consider the Bayes factor (\ref{eq:bfempi}) truncated at level $\kappa_0$
\begin{equation}
BF_{\kappa_0} =\prod_{j \: : \: l(j)\leq\kappa_0}\frac{\HG\left
(n_{j0}^{\Oneup};n_{j}^{\OneTwoup},n_{j}^{\Oneup},n_{j0}^{\OneTwoup}
\right)}{\displaystyle{\int_{0}^{\infty}}\EHG\left(n_{j0}^{\Oneup
};n_{j}^{\OneTwoup},n_{j}^{\Oneup},n_{j0}%
^{\OneTwoup},\omega_{j}\right)p(\omega_{j})d\omega_{j}}.
\end{equation}
Suppose that $\beta_\emptyset$ is as defined in Equation~\eqref
{eq:betaempty}.
If $0<\beta_\emptyset<1$ then, under $H_{0,\kappa_0}$,
\[
\lim_{n\rightarrow\infty} \log BF_{\kappa_0} =\infty
\]
and the test is consistent under the null.

\end{theorem}

\noindent\textbf{Proof.} See Appendix.
\ \rule{0.5em}{0.5em}

\begin{figure}
\includegraphics{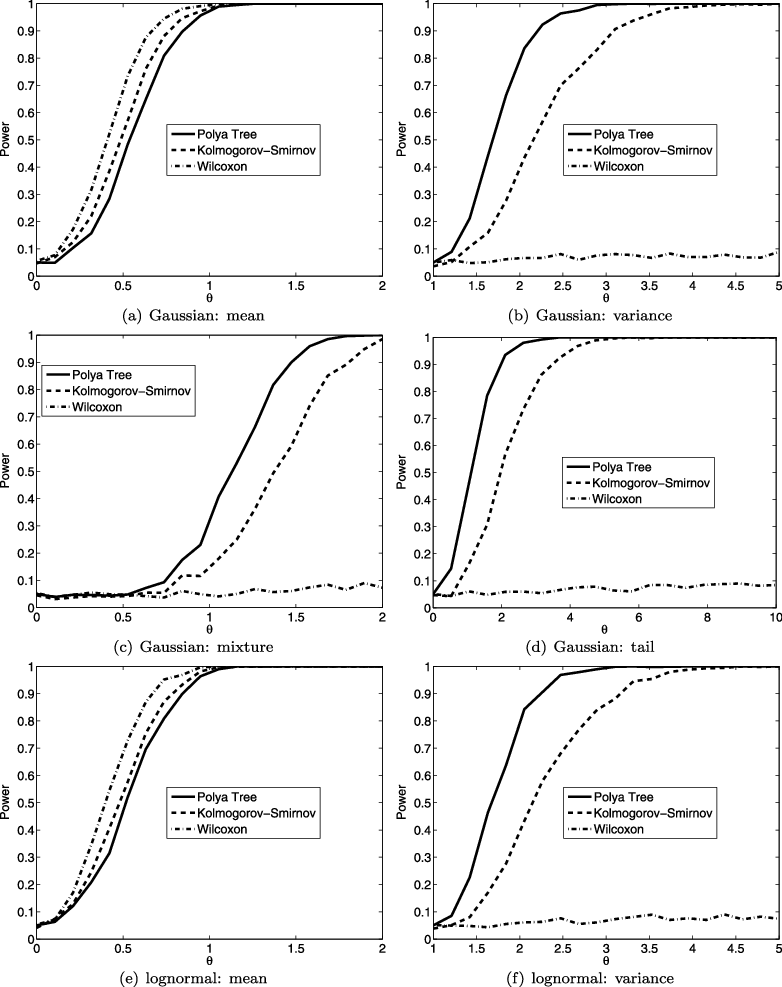}
\caption{As in Figure \ref{fig:fig1} but now using conditional Bayes
Test with $\alpha_j=m^2$.}
\label{fig:fig7}
\end{figure}

\begin{figure}
\includegraphics{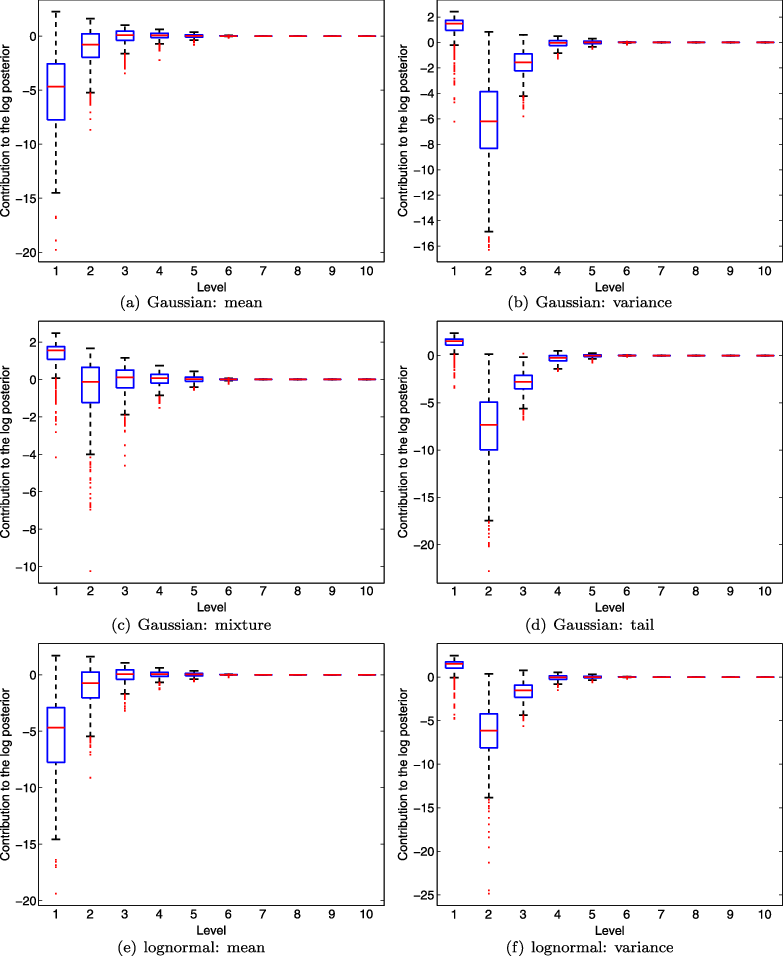}
\caption{Contribution to the Bayes Factor for different levels of the
conditional \Polya tree prior for Gaussian distribution with varying
(a) mean (b) variance (c) mixture (d) tails; log-normal distribution
with varying (e) mean (f) variance.}
\label{fig:fig9}
\end{figure}

We repeated the simulations from Section 3.3 with $\alpha= m^2$. The
results are shown in Figures~\ref{fig:fig7} and \ref{fig:fig9}. We
observe similar behaviour to the test with subjective partition but
importantly we see that the problem in detecting the difference between
normal and t-distribution is corrected. Note that no standardisation of
the data is required for this test.

\section{Sensitivity to the parameter $c$}
\label{sec:sensitivity}

The parameter $c$ acts as a precision parameter in the \Polya tree and
consequently can have an effect on the hypothesis testing procedures
previously described. In principle, the parameter can be chosen
subjectively as with precision parameters in other models (such as the
linear model). Its effect is perhaps most easily understood through the
prior variance of $P(B_{\boldsymbol{\epsilon}_k})$ which has the form
\citep{Hanson2006}
\[
\mbox{Var}\left[P(B_{\boldsymbol{\epsilon}_k})\right] = 4^{-k}
\left[\prod_{j=1}^k \frac{2cj^2+2}{2cj^2+1}-1\right].
\]
The prior variance tends to zero as $c\rightarrow\infty$ and so the
nonparametric prior places mass on distributions which more closely
resemble the centering distribution as $c$ increases. Another
consequence of this is that, under $H_1$, $c$ determines the \textit{a
priori} expected squared Euclidean distance between $F^{\Oneup
}(B_{\boldsymbol{\epsilon}_k})$ and $F^{\Twoup}(B_{\boldsymbol
{\epsilon}_k})$, where $F^{\Oneup}$ and $F^{\Twoup}$ are presumed
independently drawn from $PT(\Pi,\mathcal{A})$; this distance
diminishes as $c$ increases:
\begin{align*}
\mathbb E [(F^{(1)}(B_{\epsilon_k})-F^{(2)}(B_{\epsilon
_k}))^2]&=\mbox{Var}\left[F^{(1)}(B_{\epsilon_k})\right] + \mbox
{Var}\left[F^{(2)}(B_{\epsilon_k})\right]\\
&=4^{-k+\frac{1}{2}}
\left[\prod_{j=1}^k \frac{2cj^2+2}{2cj^2+1}-1\right].
\end{align*}

\begin{figure}
\includegraphics{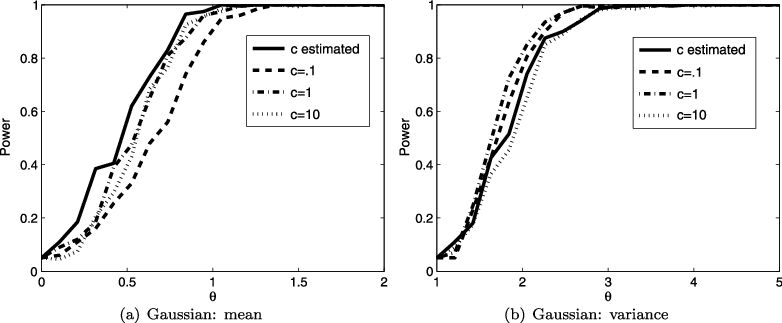}
\caption{Subjective test with empirical Bayes estimation of the
parameter $c$ for (a) mean shift (b) variance shift. Point estimates of
$c$ are obtained by maximizing both the marginal likelihood under the
null and alternative over the grid of values $10^i$ for $i=-2,-1,\ldots,3$.}
\label{fig:sensib_c_PT}
\end{figure}

\begin{figure}
\includegraphics{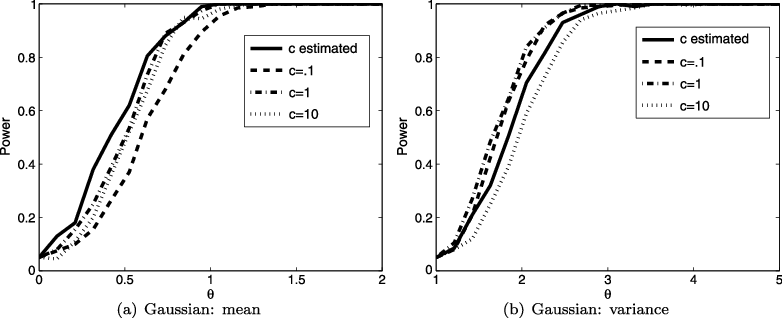}
\caption{Conditional test with empirical Bayes estimation of the
parameter $c$ for (a) mean shift (b) variance shift. Point estimate of
$c$ is obtained by maximizing the marginal likelihood under the
alternative over the grid of values $10^i$ for $i=-2,-1,\ldots,3$.}
\label{fig:sensib_c_empPT}
\end{figure}

The value of $c$ can be chosen to control the rate at which the
variances decreases. We have found values of $c$ between 1 and 10 work
well in practice. Figures~\ref{fig:sensib_c_PT} and~\ref
{fig:sensib_c_empPT} show results for different values of $c$. As with
any Bayesian testing procedure, we recommend checking the sensitivity
of their results to the chosen value of the hyperparameter $c$.

An alternative approach to the choice of $c$ in hypothesis testing is
given by \cite{Berger2001} in the context of testing a parametric
model against a nonparametric alternative. They argue that the minimum
of the Bayes factor in favour of the parametric model is useful since
the parametric model can be considered satisfactory if the ``minimum is
not too small''. It is the Bayes factor calculated using the empirical
Bayes (Type II maximum likelihood) estimate of $c$. We suggest taking a
similar approach if $c$ cannot be subjectively chosen. In the test with
subjective partition, the empirical Bayes estimates $\hat{c}$ are
calculated under $H_0$ and under $H_1$. Using these values, the Bayes
factor can be interpreted as a likelihood ratio statistic for the
comparison of the two hypotheses. In the conditional test, the
empirical Bayes estimate is calculated only under $H_1$ (since the
marginal likelihood under $H_0$ does not depend on $c$). Figures~\ref
{fig:sensib_c_PT} and~\ref{fig:sensib_c_empPT} provide results for
mean and variance shifts with $c$ estimated over a fixed grid using
this procedure.

We also performed experiments to test the sensitivity of the procedure
to the partition. Experiments (not reported here) with a partition
centered on a standard Student t distribution showed little difference
compared to a partition centered on a standard Gaussian distribution.

\section{Discussion and related work}

There have been several other approaches to testing the difference
between two distributions using \Polya tree based approaches. \cite
{Ma2011} propose the Coupling Optional \Polya tree (co-opt) prior which
extends their previous work on Optional \Polya tree priors \citep
{Wong2010}. The Optional \Polya tree defines a prior for a single
distribution. A prior is defined on the sequence of partitions used to
construct the \Polya tree. This allows the partition to be concentrated
on areas of the sample space which have the largest difference from the
base measure (which is the uniform in their case). The co-opt prior is
suitable for two distributions with a partition defined for each
distribution. The prior allows coupling of two partitions so that if a
set $A$ (which is member of the partition at level $m$) is coupled then
all subsequent partitions of $A$ will be the same for the two
distributions. This allows the posterior to concentrate these couplings
on parts of the support of the two distributions where they are similar
and so allows the borrowing of strength between the two distributions.
The prior is conjugate and can be used to both test for differences
between two distributions and to infer where these differences occur in
the posterior. The prior is particularly suited to multivariate
problems due to its ability to efficiently learn the partition of the
data. The posterior is available in closed form but can be
computationally expensive to calculate in practice with computational
time scaling exponentially with sample size.

\cite{Chen2012} propose a method for comparing $k$-samples of data
which may be censored. They test the null hypothesis that the
distribution of each sample is the same against the alternative that
the samples arise from at least two different distributions. Under the
null hypothesis, the common distribution is given a \Polya tree prior
whereas, under the alternative hypothesis, each distribution is given
an independent \Polya tree prior. All \Polya tree priors are centered
over a normal distribution whose parameters are estimated using maximum
likelihood to define an empirical Bayes procedure. Different partitions
are used for the \Polya tree distributions under the null and the
alternative hypotheses and so the partition must be truncated in order
to compute the Bayes factor, contrary to our simpler approach which
involves no truncation.

\section{Conclusions}

We have described a Bayesian nonparametric hypothesis test for real
valued data which provides an explicit measure of $\Pr(H_0 | \mathbf
{y}^{\OneTwoup})$. The test is based on a fully specified \Polya tree
prior for which we are able to derive an explicit form for the Bayes Factor.

Conditioning on a particular partition can lead to a predictive
distribution that exhibits jumps at the partition boundary points. This
is a well-known phenomenon of \Polya
tree priors and some interesting directions to mitigate its effects can
be found in~\cite{Hanson2002,Paddock2003,Hanson2006}. We do not
consider these approaches here as mixing over partitions would lose the
analytic tractability of our approach, but it is an interesting area
for future study and is considered by \cite{Chen2012}.

\vspace*{-3pt}\section*{Acknowledgements}\vspace*{-2pt}
This research was partly supported by the Oxford-Man Institute (OMI)
through a visit by FC to the OMI. FC thanks Pierre Del Moral for
helpful discussions on the proofs of consistency and acknowledges the
support of the European Commission under the Marie Curie Intra-European
Fellowship Programme. DS acknowledges the support of a Discovery Grant
from the Natural Sciences and Engineering Council of Canada. The
authors thank reviewers of earlier versions of this article for helpful
comments.



%


\section*{Appendix: Proofs}

\textbf{Proof of Theorem 1}

Clearly the log Bayes factor is
\[
\log BF_{\kappa_0}=\sum_{\{j|l(j)\leq\kappa_0\}} \log b^{(n)}_{j}.
\]
Stirling's approximation of the Gamma function allows us to write
\begin{align}
b^{(n)}_{j} \simeq&\frac{\Gamma(\alpha_{j0})\Gamma(\alpha
_{j1})}{\Gamma
(\alpha_{j0}+\alpha_{j1})}\frac{1}{\sqrt{2\pi}}\frac{\left
(\widehat{p}_{j0}%
^{\OneTwoup}\right)^{\alpha_{j0}-1/2}\left(\widehat
{p}_{j1}^{\OneTwoup}\right)^{\alpha_{j1}-1/2}%
}{\left(\widehat{p}_{j0}^{\Oneup}\right)^{\alpha_{j0}-1/2}\left
(\widehat{p}_{j1}^{\Oneup}%
\right)^{\alpha_{j1}-1/2}\left(\widehat{p}_{j0}^{\Twoup}\right
)^{\alpha_{j0}-1/2}\left(\widehat{p}%
_{j1}^{\Twoup}\right)^{\alpha_{j1}-1/2}}\nonumber\\
& \times\sqrt{\frac{n_{j}^{\Oneup}n_{j}^{\Twoup}}{n_{j}^{\OneTwoup
}}}\times
\frac{\left(\widehat{p}_{j0}^{\OneTwoup}\right)^{n_{j0}^{\OneTwoup
}}\left(\widehat{p}_{j1}%
^{\OneTwoup}\right)^{n_{j1}^{\OneTwoup}}}{\left(\widehat
{p}_{j0}^{\Oneup}\right)^{n_{j0}^{\Oneup}}\left(\widehat
{p}_{j1}^{\Oneup}\right)^{n_{j1}^{\Oneup}}\left(\widehat
{p}_{j0}^{\Twoup}\right)^{n_{j0}^{\Twoup}}%
\left(\widehat{p}_{j1}^{\Twoup}\right)^{n_{j1}^{\Twoup}}}\label
{eq:bjapprox}%
\end{align}
where
\[
\widehat{p}_{j0}^{(k)}=\frac{n_{j0}^{(k)}}{n_{j0}^{(k)}+n_{j1}^{(k)}}
\qquad\qquad\widehat{p}_{j1}^{(k)}=1-\widehat{p}_{j0}^{(k)}.
\]
We have, under the null,
\begin{equation}
\sqrt{\frac{n_{j}^{\Oneup}n_{j}^{\Twoup}}{n_{j}^{\OneTwoup
}}}\simeq\sqrt{n}\sqrt{F_0(B_j)\beta_\emptyset(1-\beta_\emptyset)}.
\label{eq:asympt1}
\end{equation}
The term
\begin{equation}
L_{j}=\frac{\left(\widehat{p}_{j0}^{\OneTwoup}\right
)^{n_{j0}^{\OneTwoup}}\left(\widehat{p}%
_{j1}^{\OneTwoup}\right)^{n_{j1}^{\OneTwoup}}}{\left(\widehat
{p}_{j0}^{\Oneup}\right)^{n_{j0}^{\Oneup}%
}\left(\widehat{p}_{j1}^{\Oneup}\right)^{n_{j1}^{\Oneup}}\left
(\widehat{p}_{j0}^{\Twoup}\right)^{n_{j0}%
^{\Twoup}}\left(\widehat{p}_{j1}^{\Twoup}\right)^{n_{j1}^{\Twoup
}}}%
\label{eq:ellj}
\end{equation}
is a likelihood ratio for testing composite hypotheses
\[
H_{j0}:p_{j0}^{\Oneup}=p_{j0}^{\Twoup}=p_{j0}^{\OneTwoup}\text{ vs
}H_{j1}:\left(p_{j0}%
^{\Oneup},p_{j0}^{\Twoup}\right)\in\lbrack0,1]^{2}%
\]
with $n_{j0}^{\Oneup}\sim\Bino\left(n_{j}^{\Oneup},p_{j0}^{\Oneup
}\right)$ and $n_{j0}^{\Twoup}\sim
\Bino\left(n_{j}^{\Twoup},p_{j0}^{\Twoup}\right)$. Clearly
$\widehat{p}_{j0}^{\OneTwoup}$ and\break $\left(\widehat
{p}_{j0}^{\Oneup}, \widehat{p}_{j0}^{\Twoup}\right)$ are the
maximum likelihood estimators under
$H_{j0}$ and $H_{j1}$ respectively. It follows that, under $H_{j0}$,
$-2\log L_{j}$ asymptotically follows a $\chi^{2}$ distribution \citep
{Wilks1938}.

Finally, if $\beta_\emptyset(1-\beta_\emptyset)>0$ and using
Equation~\eqref{eq:asympt1}, then Theorem 1 follows.

\noindent\textbf{Proof of Theorem 2}

If $F^{\Oneup}(B_j)=0$ or $F^{\Twoup}(B_j)=0$, then we have trivially
$\log(b_j)=0$. We assume that
\begin{eqnarray}
F^{\Oneup}(B_j)F^{\Twoup}(B_j)>0\label{eq:assumptions1}~\\
0<\beta_\emptyset<1.
\label{eq:assumptions2}
\end{eqnarray}
If $p_{j0}^{\Oneup}=p_{j0}^{\Twoup}$ then, from the previous section,
$\log(b_j)$ goes to $\infty$ in $o(\log(n))$. We consider now the
case $p_{j0}^{\Oneup}\neq p_{j0}^{\Twoup}$.

Let $\widehat{\beta}^{(n)}_{j}=n_{j}^{\Oneup}/n_{j}^{\OneTwoup}$,
\begin{align*}
\beta_j&=\lim_{n\rightarrow\infty}\widehat{\beta}^{(n)}_{j}=\frac
{\beta_\emptyset F^{\Oneup}(B_j)}{F^{\OneTwoup}(B_j)}
\end{align*}
with $F^{\OneTwoup}(B_j)=\beta_\emptyset F^{\Oneup}(B_j)+(1-\beta
_\emptyset) F^{\Twoup}(B_j)$. Under assumptions (\ref
{eq:assumptions1}) and (\ref{eq:assumptions2}), we have $0<\beta
_j<1$. Let $L_{j}$ be
defined as in Equation \eqref{eq:ellj}. We have
\[
\log L_{j} = \eta_j - n_j^{\OneTwoup} \zeta_j
\]
where
\[
\eta_j = n_{j}^{\OneTwoup}H_{p_{j0}^{\OneTwoup}}\left(\widehat
{p}_{j0}^{\OneTwoup}
\right)-n_{j}^{\Oneup}H_{p_{j0}^{\Oneup}}\left(\widehat
{p}_{j0}^{\Oneup}\right)-n_{j}^{\Twoup}
H_{p_{j0}^{\Twoup}}\left(\widehat{p}_{j0}^{\Twoup}\right),
\]
\[
\zeta_j = \widehat{\beta}^{(n)}_{j}H_{1}\left(p_{j0}^{\Oneup
}\right)+\left(1-\widehat{\beta}^{(n)}_{j}\right)H_{1}
\left(p_{j0}^{\Twoup}\right)-H_{1}\left(p_{j0}^{\OneTwoup}\right),
\]
\[
p_{j0}^{(k)}=\lim_{n\rightarrow\infty}\widehat{p}_{j0}^{(k)},\text
{ }k=1,2,\{1,2\}
\]
and the function $H_{p}:x\in(0,1)\rightarrow\mathbb{R}_+$ is defined
for $p\in(0,1)$ by
\[
H_{p}(x)=x\log\left(\frac{x}{p}\right)+(1-x)\log\left(\frac
{1-x}{1-p}\right).
\]
Consider first the term $\zeta_j$. We have
\[
p_{j0}^{\OneTwoup}=\beta_j p_{j0}^{\Oneup}+(1-\beta
_j)p_{j0}^{\Twoup}
\]
and $n\longrightarrow\infty$, therefore
\[
\zeta_j\longrightarrow\beta_{j}H_{1}\left(p_{j0}^{\Oneup}\right
)+\left(1-\beta_{j}\right)H_{1}
\left(p_{j0}^{\Twoup}\right)-H_{1}\left(p_{j0}^{\OneTwoup}\right).
\]
As the function $H_p$ is convex, it follows that $\zeta_j$ tends to a
positive constant
if $p_{j0}^{\Oneup}\neq p_{j0}^{\Twoup}$. Hence $n_{j}^{\OneTwoup
}\zeta_j$
tends to $\infty$ in $o\left(n\right)$.

Consider now $\eta_j$ which is approximately equal to
\begin{align*}
Y_{j}=&\frac{n_{j}^{\OneTwoup}}{2p_{j0}^{\OneTwoup}\left
(1-p_{j0}^{\OneTwoup}\right)}\left(\widehat{p}%
_{j0}^{\OneTwoup}-p_{j0}^{\OneTwoup}\right)^{2}-\frac{n_{j}^{\Oneup
}}{2p_{j0}^{\Oneup}%
\left(1-p_{j0}^{\Oneup}\right)}\left(\widehat{p}_{j0}^{\Oneup
}-p_{j0}^{\Oneup}\right)^{2}\\
&-\frac{n_{j}^{\Twoup}%
}{2p_{j0}^{\Twoup}\left(1-p_{j0}^{\Twoup}\right)}\left(\widehat
{p}_{j0}^{\Twoup}-p_{j0}^{\Twoup}\right)^{2}.
\end{align*}
Then
\begin{align*}
Y_{j} =&-\frac{n_{j}^{\OneTwoup}\widehat{\beta}^{(n)}_{j}\left
(1-\widehat{\beta}^{(n)}_{j}\right)}{2p_{j0}^{\OneTwoup}%
\left(1-p_{j0}^{\OneTwoup}\right)}\left[\widehat{p}_{j0}^{\Oneup
}-\widehat{p}_{j0}^{\Twoup}%
-\left(p_{j0}^{\Oneup}-p_{j0}^{\Twoup}\right)\right]^{2}\\
& +\frac{n_{j}^{\OneTwoup}\widehat{\beta}^{(n)}_{j}\left(1-\rho
_{j}^{\Oneup}\right)}{2p_{j0}^{\OneTwoup}
\left(1-p_{j0}^{\OneTwoup}\right)}\left(\widehat{p}_{j0}^{\Oneup
}-p_{j0}^{\Oneup}\right)^{2}
+\frac{n_{j}^{\OneTwoup}\left(1-\widehat{\beta}^{(n)}_{j}\right
)\left(1-\rho_{j}^{\Twoup}\right)}{2p_{j0}^{\OneTwoup}%
\left(1-p_{j0}^{\OneTwoup}\right)}\left(\widehat{p}_{j0}^{\Twoup
}-p_{j0}^{\Twoup}\right)^{2}%
\end{align*}
where
\[
\rho_{j}^{(k)}=\frac{p_{j0}^{\OneTwoup}\left(1-p_{j0}^{\OneTwoup
}\right)}{p_{j0}^{(k)}%
\left(1-p_{j0}^{(k)}\right)}.
\]
We have asymptotically
\begin{align*}
Y_{j} \simeq& \; n F^{\OneTwoup}(B_j)\left(\frac{-\beta_{j}\left
(1-\beta_{j}\right)}{2p_{j0}^{\OneTwoup}%
\left(1-p_{j0}^{\OneTwoup}\right)}\left[\widehat{p}_{j0}^{\Oneup
}-\widehat{p}_{j0}^{\Twoup}%
-\left(p_{j0}^{\Oneup}-p_{j0}^{\Twoup}\right)\right]^{2}\right.\\
& \left. +\frac{\beta_{j}\left(1-\rho_{j}^{\Oneup}\right
)}{2p_{j0}^{\OneTwoup}
\left(1-p_{j0}^{\OneTwoup}\right)}\left(\widehat{p}_{j0}^{\Oneup
}-p_{j0}^{\Oneup}\right)^{2}
+\frac{\left(1-\beta_{j}\right)\left(1-\rho_{j}^{\Twoup}\right
)}{2p_{j0}^{\OneTwoup}%
\left(1-p_{j0}^{\OneTwoup}\right)}\left(\widehat{p}_{j0}^{\Twoup
}-p_{j0}^{\Twoup}\right)^{2}\right).
\end{align*}
We have a quadratic form
\[
Y_{j}\simeq - n F^{\OneTwoup}(B_j)\frac{\beta_{j}\left(1-\beta
_{j}\right)}{2p_{j0}^{\OneTwoup}%
\left(1-p_{j0}^{\OneTwoup}\right)}\left[
\begin{array}
[c]{cc}
\left(\widehat{p}_{j0}^{\Oneup}-p_{j0}^{\Oneup}\right) &\ \left
(\widehat{p}_{j0}^{\Twoup}-p_{j0}^{\Twoup}\right)
\end{array}
\right] A\left[
\begin{array}
[c]{c}%
\left(\widehat{p}_{j0}^{\Oneup}-p_{j0}^{\Oneup}\right)\\
\left(\widehat{p}_{j0}^{\Twoup}-p_{j0}^{\Twoup}\right)
\end{array}
\right]
\]
with square matrix
\[
A=\left[
\begin{array}
[c]{cc}%
\dfrac{\rho_{j}^{\Oneup}-\beta_{j}}{1-\beta_{j}} & -1\\
-1 & \dfrac{\rho_{j}^{\Twoup}-1+\beta_{j}}{\beta_{j}}%
\end{array}
\right]
\]
and so asymptotically
\[
\sqrt{n}\left[
\begin{array}
[c]{c}%
\left(\widehat{p}_{j0}^{\Oneup}-p_{j0}^{\Oneup}\right)\\
\left(\widehat{p}_{j0}^{\Twoup}-p_{j0}^{\Twoup}\right)
\end{array}
\right] \sim\mathcal{N}\left( 0_{2},\left(
\begin{array}
[c]{cc}
\dfrac{p_{j0}^{\Oneup}\left(1-p_{j0}^{\Oneup}\right)}{\beta
_{j}F^{\OneTwoup}(B_j)} & 0\\
0 & \dfrac{p_{j0}^{\Twoup}\left(1-p_{j0}^{\Twoup}\right)}{(1-\beta
_{j})F^{\OneTwoup}(B_j)}%
\end{array}
\right) \right)
\]
which is independent of the sample size, and it follows that $Y_j$
asymptotically follows a scaled $\chi^2$ distribution and $\log L_j$
goes to $-\infty$ in $o(n)$. To conclude, we have that in
Equation~\eqref{eq:bjapprox}
\begin{equation*}
\sqrt{\frac{n_{j}^{\Oneup}n_{j}^{\Twoup}}{n_{j}^{\OneTwoup
}}}\simeq\sqrt{n}\sqrt{F^{\OneTwoup}(B_j)\beta_j(1-\beta_j)}.
\label{eq:asympt2}
\end{equation*}

It follows that
\begin{itemize}
\item If $F^{\Oneup}(B_j)F^{\Twoup}(B_j)=0$, $\log(b_j)=0$.
\item If $F^{\Oneup}(B_j)F^{\Twoup}(B_j)>0$
\begin{itemize}
\item If $p_{j0}^{\Oneup}\neq p_{j0}^{\Twoup},$ the log-contribution
$\log(b_{j})$ goes to
$-\infty$ at a rate of $o(n)$.
\item If $p_{j0}^{\Oneup}=p_{j0}^{\Twoup}$, the log-contribution
$\log(b_{j})$ goes to $+\infty$ at a rate of $o(\log n)$.
\end{itemize}
\end{itemize}

\noindent\textbf{Proof of Theorem 3}

The condition $\beta_\emptyset(1-\beta_\emptyset)$ implies that
$\beta_j(1-\beta_j)>0$ for all $j$. Let
\begin{equation}
b^{(n)}_{j}=\frac{\HG\left(n_{j0}^{\Oneup};n_{j}^{\OneTwoup
},n_{j}^{\Oneup},n_{j0}^{\OneTwoup}\right)}%
{\displaystyle{\int_{0}^{\infty}}\exp\left\{ u(\omega_j)\right\}
d\omega_{j}}%
\label{eq:bj}
\end{equation}
where $u(\omega_{j})=\log\EHG\left(n_{j0}^{\Oneup
};n_{j}^{\OneTwoup},n_{j}^{\Oneup},n_{j0}%
^{\OneTwoup},\omega_{j}\right)+ \log p(\omega_{j})$. Then
\[
\log BF_{\kappa_0}=\sum_{\{j|l(j)\leq\kappa_0\}} \log b^{(n)}_{j}.
\]
Under the conditions $\beta_j(1-\beta_j)>0$, the maximum likelihood
estimate $\widehat{\omega}_j$ of the parameter $\omega_j$ in the
extended hypergeometric distribution converges in probability to the
true parameter~\citep[p. 944]{Harkness1965}. We can therefore use a
Laplace approximation~\citep{Kass1995} of the
denominator in \eqref{eq:bj}, we obtain for $n_{j0}^{\Oneup}$ large
\[
b^{(n)}_{j}\simeq\frac{\HG\left(n_{j0}^{\Oneup};n_{j}^{\OneTwoup
},n_{j}^{\Oneup},n_{j0}^{\OneTwoup}%
\right)}{\sqrt{2\pi}|\widehat{\Sigma}_{j}|^{1/2}\EHG\left
(n_{j0}^{\Oneup};n_{j}^{\OneTwoup},n_{j}%
^{\Oneup},n_{j0}^{\OneTwoup},\widehat{\omega}_{j}\right)p\left
(\widehat{\omega}_{j}\right)}%
\]
where $\widehat{\omega}_{j}=\text{argmax}_{\omega_j} u(\omega_j) $
and $\widehat{\Sigma}_{j}%
^{-1}=-\mathbf{D}^{2}u_{j}(\widehat{\omega}_j)$, where $\mathbf
{D}^{2}u_{j}(\widehat{\omega}_j)$ is the Hessian matrix of second
derivatives. The ratio%
\[
r_{j}=\frac{\HG\left(n_{j0}^{\Oneup};n_{j}^{\OneTwoup
},n_{j}^{\Oneup},n_{j0}^{\OneTwoup}%
\right)}{\EHG\left(n_{j0}^{\Oneup};n_{j}^{\OneTwoup},n_{j}^{\Oneup
},n_{j0}^{\OneTwoup},\widehat{\omega}%
_{j}\right)}=\frac{\EHG\left(n_{j0}^{\Oneup};n_{j}^{\OneTwoup
},n_{j}^{\Oneup},n_{j0}^{\OneTwoup}%
,1\right)}{\EHG\left(n_{j0}^{\Oneup};n_{j}^{\OneTwoup
},n_{j}^{\Oneup},n_{j0}^{\OneTwoup},\widehat{\omega
}_{j}\right)}%
\]
is a likelihood ratio for testing the composite hypotheses%
\[
H_{j0}:\omega_{j}=1\text{ vs }H_{j1}:\omega_{j}>0,
\]
hence $-2\log r_{j}$ is asymptotically $\chi^{2}$-distributed \citep
{Wilks1938}.
And as $|\widehat{\Sigma}_{j}|\rightarrow0$ as $n \rightarrow\infty
$, then $b^{(n)}_{j}\rightarrow\infty$ for all $j$ as $n\rightarrow
\infty$.

\end{document}